\documentclass[fleqn,11pt]{wlscirep}

\usepackage{mathtools}
\usepackage{bm}
\usepackage{isomath}
\usepackage[displaymath,mathlines]{lineno}

\title{Extracting quasi-steady Lagrangian transport patterns from the ocean circulation: An application to the Gulf of Mexico}

\author[1,*]{R.\ Duran}
\author[2]{F.\ J.\ Beron-Vera}
\author[3]{M.\  J.\ Olascoaga}

\affil[1]{College of Earth Ocean and Atmospheric Sciences, Oregon State University, Corvallis, Oregon, USA}
\affil[*]{Corresponding author: rduran@ceoas.oregonstate.edu}
\affil[2]{Department of Atmospheric Sciences, Rosenstiel School of Marine and Atmospheric Science, University of Miami, Miami, Florida, USA} 
\affil[3]{Department of Ocean Sciences, Rosenstiel School of Marine and Atmospheric Science, University of Miami, Miami, Florida, USA}

\begin{abstract}
We construct a climatology of Lagrangian coherent structures (LCSs), the concealed skeleton that shapes transport, with a twelve-year-long data-assimilative simulation of the sea-surface circulation in the Gulf of Mexico (GoM). Computed as time-mean Cauchy-Green strain tensorlines of the climatological velocity, the climatological LCSs (cLCSs) unveil recurrent Lagrangian circulation patterns. cLCSs strongly constrain the ensemble-mean Lagrangian circulation of the instantaneous model velocity, thus we show that a climatological velocity may preserve meaningful transport information. Also, the climatological transport patterns we report agree well with GoM kinematics and dynamics, as described in several previous observational and numerical studies.  For example, cLCSs identify regions of persistent isolation, and suggest that coastal regions previously identified as high-risk for pollution impact, are regions of maximal attraction. Also, we show examples where cLCSs are remarkably similar to transport patterns observed during the \emph{Deepwater Horizon} and \emph{Ixtoc} oil spills, and during the Grand LAgrangian Deployment (GLAD) experiment.  Thus, it is shown that cLCSs are an efficient way of synthesizing vast amounts of Lagrangian information. The cLCS method confirms previous GoM studies, and contributes to our understanding by revealing the persistent nature of the dynamics and kinematics treated therein.
\end{abstract}

\begin{document}
\flushbottom
\maketitle
\thispagestyle{empty}

\section*{Introduction.}
\noindent
Lagrangian transport is a difficult oceanographic problem for which solutions are frequently needed.  Sensitivity to initial conditions
or to the precision of the velocity field require more attention to detail than what we are usually able to afford. Even if those details could be resolved, making progress with only a few general guidelines would be ideal, appraising many new applications as practical. For example, those tasked with planning for environmental-pollution response and prevention would like to have information applicable to generic oil spills: Identification of recurrent trajectory patterns at the sea surface, identification of regions with a higher risk of contamination, and identification of isolated areas, where pollution is unlikely to enter or leave. 
We therefore ask if it is possible to find structures that tend to organize transport, and that evolve slowly relative to a typical Lagrangian timescale.  For instance, the Lagrangian-velocity decorrelation time, which is order one day at the sea surface, and that constitutes an indicator of Lagrangian predictability\cite{LaCasce2008}.\\
\noindent
In this paper, we analyze a long record of surface currents in the Gulf of Mexico (GoM) from a data-assimilative simulation
using results from nonlinear dynamical systems theory, that enable the objective (observer-independent) identification of key material lines that organize Lagrangian transport, often referred to as Lagrangian coherent structures (LCSs) \cite{Haller-Yuan-00}. Samelson\cite{Samelson2013} reviews the original heuristic approaches and includes a fluid-dynamical account of the terminology;  Haller \cite{Haller2015} reviews the recent, rigorous theory employed here.\\
\noindent
We show that it is possible to construct a climatology of LCSs that strongly constrains the ensemble-mean instantaneous ocean-model Lagrangian circulation. The climatological transport patterns also agree well with kinematics and dynamics described in several previous studies e.g. by identifying regions of persistent isolation, or by suggesting that the coastal regions that have been reported as high-risk for pollution impact are regions of maximal attraction. Finally, climatological LCSs agree remarkably well with surface-drifter and oil-spill distributions and beachings.

\section*{Methods.}
\subsection*{LCS.}
Let $\mathbf{v}(\mathbf{x},t)$ be a two-dimensional velocity
field, where $\mathbf{x}\in U \subset \mathbb{R}^2$ denotes position and $t\in [a,b] \subset \mathbb{R}$ is time; different time intervals define different finite-time dynamical systems. For each $\mathbf{x}_0 \in U$, let $\mathbf{F}^t_{t_0}(\mathbf{x}_0): \mathbf{x}_0 \mapsto \mathbf{x}(t;\mathbf{x}_0,t_0) $ be the flow map that associates times $t_0$ and $t$ positions of fluid particles, which evolve according to
\begin{eqnarray}
  \frac{d\mathbf{x}}{dt}  = \mathbf{v}(\mathbf{x},t).  
  \label{eq:dxdt}
\end{eqnarray} 
A commonly used, objective measure of material
deformation is the right Cauchy--Green (CG) strain tensor, given by
\begin{eqnarray}
  \mathbf{C}^t_{t_0}(\mathbf{x}_0) \coloneqq
   \left[\mathrm{D} \mathbf{F}^t_{t_0}(\mathbf{x}_0)\right]^\top 
  \mathrm{D} \mathbf{F}^t_{t_0}(\mathbf{x}_0)
  \label{eq:C}
\end{eqnarray}
where
\begin{eqnarray}
\mathrm{D} \mathbf{F}^t_{t_0}(\mathbf{x}_0) =
 \begin{bmatrix}
   \dfrac{\partial x}{\partial x_0} & \dfrac{\partial x}{\partial y_0}\\[0.16 in]
  \dfrac{\partial y}{\partial x_0} & \dfrac{\partial y}{\partial y_0}\\
 \end{bmatrix}
 \label{eq:DF}
\end{eqnarray}

\noindent
In the above definitions, both $t<t_0$ or $t>t_0$ are acceptable. \\

\noindent
Let $0 < \lambda_1(\mathbf{x}_0) < \lambda_2(\mathbf{x}_0)$ and $\boldsymbol{\xi}_1(\mathbf{x}_0) \perp \boldsymbol{\xi}_2(\mathbf{x}_0)$ be eigenvalues and normalized eigenvectors of
\eqref{eq:C}.  
A local normal-growth measure of the unit normal, $\mathbf{n}_0$, along a
material line at time $t_0$ is given by \cite{Haller-11}:
\begin{eqnarray}
  \rho_{t_0}^t(\mathbf{x}_0,\mathbf{n}_0) \coloneqq  \frac{1}{\sqrt{\smash[b]{\mathbf{n}_0
  \cdot \mathbf{C}_{t_0}^t(\mathbf{x}_0)^{-1}\mathbf{n}_0}}}. 
  \label{eq:rho}
\end{eqnarray}
The LCSs are especial material lines that shape global (i.e., over $U$) Lagrangian
transport patterns produced by $\mathbf{v}(\mathbf{x},t)$.  Of particular interest
 for our purposes are attracting LCSs, as these delineate Lagrangian
transport pathways.  An LCS that attracts nearby particle trajectories
over a finite-time interval $[t,t_0]$, where $t = t_0 + T$ and $T<0$, is a squeezing
Cauchy--Green strain tensorline or \emph{squeezeline} \cite{Olascoaga-etal-13}, i.e., a curve
$s \mapsto \mathbf{x}(s)$ which satisfies \cite{Haller-Beron-12,Farazmand-etal-14,Haller2015}
\begin{eqnarray}
  \frac{d\mathbf{x}}{ds} = \boldsymbol{\xi}_1(\mathbf{x})
  \label{eq:sq1}
\end{eqnarray}
and \cite{Beron-etal-15}
\begin{eqnarray}
  \rho_{t_0}^t(\mathbf{x}) = \sqrt{\lambda_2(\mathbf{x})} > 1.
  \label{eq:sq2}
\end{eqnarray}
If the velocity $\mathbf{v}$ is non-divergent, then \eqref{eq:sq2} is guaranteed to be satisfied. The most attracting LCSs in forward time are those with the largest back-in-time normal repulsion  $\rho_{t_0}^t(\mathbf{x})$. This implies tangential stretching in forward time for a nondivergent flow. For simplicity we no longer write $\rho$'s dependence on $t_0,t$ and $\mathbf{x}$ below.

\subsection*{Velocity data, numerics and timescale.}
\noindent
For the velocity $\mathbf{v}(\mathbf{x},t)$ we use 12 years of daily sea-surface velocity from the Hybrid-Coordinate Ocean Model\cite{Bleck-02} (HyCOM), forced by the US Navy Operational Global Atmospheric Prediction System. The resolution is about 4 km,  adequate for our search of persistent material deformation shaping global transport. In the Supporting Information (Appendix A), we show that the motions we report are caused by confluence (divergence-free attraction) rather than convergence (attraction with negative divergence). \\
\noindent
From the HyCOM-GOM10.04 analysis, experiment 20.1 was used for years 2003 through 2009, experiment 31 for January through March 2010, and experiment 32.5  for April 2010 through year 2014. This 12-year period (2003-2014) was used because HyCOM simulations include the Navy Coupled Ocean Data Assimilation\cite{Cummings-05,Cummings-Smedstad-13} (NCODA).\\
\noindent For simplicity, each data-year is defined to be the first 360 days of the calendar year, and therefore months are composed of 30 days. The only exception is 2003 spanning days 2 to 361 due to availability. Temporal resolution is for the most part a daily instantaneous field except for a 4-day gap in 2004, a 2-day gap in 2009 and a 1.5-day gap in 2014. Cubic interpolation was used to remediate these gaps and keep the time between velocity fields at 24 hours.  The first day of the climatology is obtained by averaging the first day of the 2003-2014 time series, and so on.\\
\noindent
We note that the kinetic energy spectra as represented by models at resolutions of up to about 1 km are steep enough for bulk Lagrangian calculations to be largely insensitive to fine velocity details in space and time \cite{Beron-LaCasce-16,Beron-Olascoaga-09}. Additionally, Keating et al. \cite{Keating-etal-11} showed that interpolation can ameliorate even a very coarse temporal resolution. \\ 
\noindent
All integrations are done with a Runge--Kutta 4(5) pair (i.e. with adaptative time step) and cubic interpolations. The computational domain covers the GoM with a mean grid spacing of 1.7 km; an auxiliary grid of 4 points separated by 0.1 km and centered at each grid point of the main grid is used to evaluate the centered derivatives with which \eqref{eq:DF} is approximated. 
The numerical implementation of geodesic LCS detection is documented at length \cite{Haller-Beron-12,Olascoaga-etal-13,Hadjighasem-etal-13,Farazmand-etal-14}; a software tool is also available \cite{Onu-etal-15}.\\
\noindent
A few days to a week is a critical timescale for oil-spill response \cite{API-NOAA-USCG-EPA2010},  search and rescue operations \cite{Melsom2012,Chen2012}, and larval recruitment and algal blooms \cite{Lalli-Parsons1993,Miller2004}. Thus, we choose $T=-7$ days.

\subsection*{Climatological LCSs.}
\noindent
Using each month's 30 days of data, we compute CG tensors for the dynamical systems with $T=-7$ days fixed, and initial times taken from the set $t_0\in\{8,10,12,\dots,30\}$ days. The monthly-mean CG tensor is then the average of the CG tensors from these back-in-time, 7-day flow maps, initiated every other day in that month. We refer to LCSs computed from the monthly-mean CG tensor as \emph{climatological LCSs} (cLCSs).  We compute the corresponding average repulsion rate by diagonalizing the average CG tensor. We refer to this average as climatological attraction or (due to our velocity being very nearly nondivergent) climatological  stretching, and denote it c$\rho$.
\noindent

\noindent
 A welcome finding through sensitivity tests was that our results are robust. Tests using $T \in \{-5, -7, -10, -15, -20\}$ days, suggest that our results do not depend sensitively on $T$. Tests using instantaneous six-, twelve-, or 24-hourly velocity fields suggest that different temporal resolutions will not affect the results.  Finally, we also get the same results when using a second order Runge-Kutta.

\subsection*{Additional data.}
\noindent
Beyond comparisons with ensembles using  the instantaneous model velocity, and with previous studies, we use three data sets to evaluate results from the LCS climatology.
First we use surface oil images during the \emph{Deepwater Horizon} spill (DwH), produced by the NOAA Experimental
Marine Pollution Surveillance Reports (EMPSR; http://\allowbreak
www.ssd.noaa.gov/\allowbreak PS/MPS/\allowbreak deepwater.html).
They delineate surface oil using satellite imagery
from active and passive sensors, overflights and in situ observations. Further support for our results is provided by surface-drifter trajectory
data from the Grand LAgrangian Deployment (GLAD) in the
northern GoM \cite{Olascoaga-etal-13,Poje-etal-14,Jacobs-etal-14,Coelho-etal-15}. Finally, we use data from the \emph{Ixtoc} spill (1979-1980). Availability of observations with good temporal and spatial coverage for this event is restricted to data from the Coastal Zone Color Scanner (CZCS, Nimbus-7 satellite) and Landsat Multispectral Scanner (MSS, Landsat 1-5 satellites).  These data were acquired and carefully processesed by Sun et al. \cite{Sun2015}; we also use the beaching locations in their table 1.

\section*{Results and discussion.}
\subsection*{Comparison with the ensemble-mean transport produced by the model.}
\label{inst}

\noindent
We first establish a relationship between the cLCSs and transport supported by the instantaneous velocity used to compute the velocity climatology. Consider an arbitrary tracer distribution $f_0(\mathbf{x}_0)$ at time $t_0$.  The image of a point $\mathbf{x}_0$ under the flow at time $t = t_0+T$, is $\mathbf{x}_{t_0+T} = \mathbf{F}_{t_0}^{t_0+T}(\mathbf{x}_0)$. The advected image of a conserved tracer distribution is (e.g. Froyland et al. \cite{Froyland-etal-07}):
\begin{eqnarray}
 f_{t_0+T}(\mathbf{x}_{t_0+T}) = f_0\left(\mathbf{F}^{t_0}_{t_0+T}(\mathbf{x}_{t_0+T})\right) \det \mathrm{D} \mathbf{F}^{t_0}_{t_0+T}(\mathbf{x}_{t_0+T})
 \label{eq:pushfwd}
\end{eqnarray}

\begin{figure}[h]
  \includegraphics[width=\textwidth]{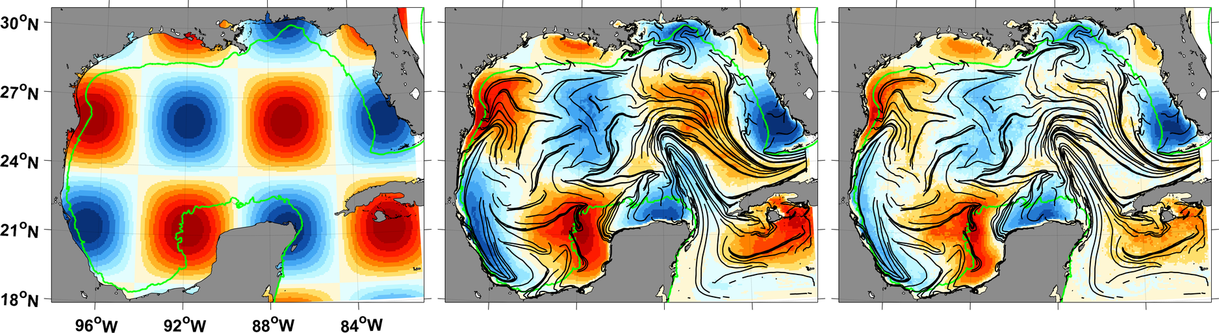}
  \caption{The left panel is the initial distribution (colors, arbitrary units) of a tracer, the 50-m isobath is plotted in green in all panels. The middle panel shows the ensemble-mean distribution resulting from 7-day advections of the initial distribution, under the instantaneous velocity for each January in the climatology, plotted over January's cLCSs (black lines). The right panel is the same as the middle panel except that the ensemble-mean distribution is the result of 14-day advections. The arbitrary colorscale limits are the same in all three panels.  }
  \label{fig:fig04}
\end{figure}

\noindent
(Note that flow maps preserve orientation, and therefore the Jacobian determinant cannot be negative.)
Consider an ensemble of backward trajectories $\{\mathbf{F}_{t_0 + T}^{t_0}(\mathbf{x}_{t_0 + T})\}$ with different $t_0$ for each $\mathbf{x}_{t_0+T}$. The corresponding ensemble-mean backward flow map $\mathbf{\overline{F}}_{T}^{\, 0}(\mathbf{x}_T)$, can be used to evaluate \eqref{eq:pushfwd}, producing an ensemble-mean tracer distribution $\overline{f}_T(\mathbf{x}_T)$.\\
\noindent
Let $f_0(\mathbf{x}_0) = \sin \frac{x_0}{2} \sin \frac{y_0}{2}$ with $\mathbf{x}_0 = 0$ roughly at the center of the GoM, conveniently chosen to facilitate the visualization of advection patterns. If we take $t_0$ over January along each of the 12 years of simulation, the ensemble-mean distributions $\overline{f}_T(\mathbf{x}_T)$ at $T=7$ and  $T=14$ days show that the instantaneous circulation is strongly constrained by the corresponding January cLCSs (Fig.\ \ref{fig:fig04}). LCSs are by construction material lines, despite cLCSs not preserving this property, the tracer distribution often stretches along cLCSs, e.g. meridionally along the western GoM, just north of the Yucatan shelf and along the 50-m isobath on the western Yucatan shelf. Also, cLCSs seem to serve as barriers e.g. within the Yucatan shelf around 91$^\circ$W, separating blue and red tracer. These results do not depend on the choice of month.

\subsection*{Comparison with known transport patterns.}
\begin{figure}[h]
  \includegraphics[width=\textwidth]{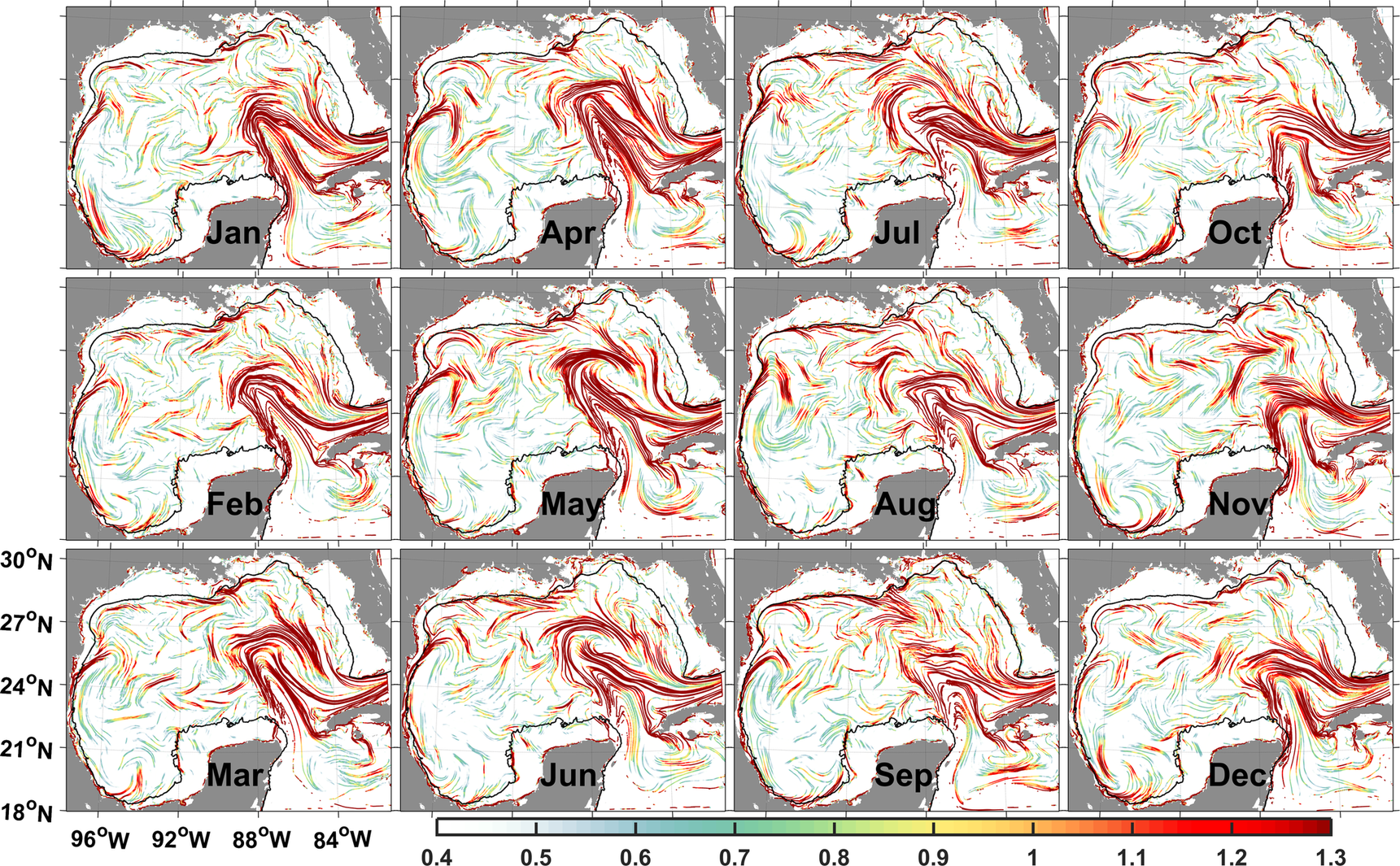}
  \caption{Monthly climatological LCSs (each column is roughly a season) colored according to their climatological attraction strength  $\ln \mathrm{c}\rho$ .  The 50-m isobath is indicated in black.  }
  \label{fig:fig01a}
\end{figure}

\noindent
A characterization highlighting important kinematical features in the GoM emerges from the monthly cLCS maps (Fig. \ref{fig:fig01a}).  We illustrate this through comparisons with known transport patterns and, in a subsection below, with observations.\\

\subsubsection*{The Loop Current.}

\noindent
The Loop Current (LC), a region of persistent attraction, is the predominant feature of the cLCS fields in the GoM interior.  
Relative to previous months, the LC outline does not reach as north in the last three months of the year.  This is consistent with different multi-year studies showing that, in any given year, eddies will most likely have shed by the time the first 9 months of that year are over, cf.\ Fig.\ 12 of Vukovich\cite{Vukovich2007} and Fig.\ 4 of Lindo-Atichati et al. \cite{Lindo-Atichati2013}\\

\subsubsection*{Western GoM.}

\noindent
Between 92-96$^\circ$W and 19-23$^\circ$N, climatological attraction c$\rho$ is often weak suggesting stagnation relative to  the rest of the GoM interior (i.e. deeper than 50m). However, along the western-boundary 50-m isobath, meridional  stretching tends to be relatively strong. Consistently, northward advection along the western margin and retention within the southwestern GoM are the two characteristic dispersion scenarios identified by Zavala-Sanson et al.\cite{Zavala-Sanson-etal-17} from drifters released in years 2007-2014, between 93-96$^\circ$W and 19-20.5$^\circ$N. \\
Our climatology often shows offshore transport originating from the western-boundary 50-m isobath around 24-26$^\circ$N, and from the southwestern 50-m isobath around 19$^\circ$N and 92-94$^\circ$W. Enhanced cross-shelf transport in both these regions was identified through satellite data between 1997-2007 by {Mart{\'{\i}}nez-L{\'o}pez} \& {Zavala-Hidalgo} \cite{ML-ZH-2009} (cf. their Fig. 7); they propose that both cases of offshore transport are due to convergence of alongshore wind-driven currents, with the western case (24-26 $^\circ$N) modulated by the presence of cyclonic and anticyclonic eddies. Zhang \& Hetland \cite{Zhang-Hetland-2012} reached a similar conclusion over the Louisiana-Texas (La-Tex) shelf. The study by Gough et al.\cite{Goughetal2017} is, to our knowledge, the first to apply the techniques described in this paper. They confirm offshore transport around 24-26$^\circ$N by computing cLCSs from an 18-year simulation using the Nucleus for European Modelling of the Ocean (NEMO) model. Their cLCSs agree well with transport from historical drifter observations and synthetic drifters advected with their instantaneous velocity.    \\

\subsubsection*{Coastal risk.}
Coastal cLCSs imply an increased risk of environmental impact to the nearby coastline: Lagrangian parcels are persistently attracted to where their trajectories  may be subject to effective cross-shelf drivers such as Stokes drift \cite{LeHenaff2012,Weisberg2017}. \\
\noindent
The vicinity of the Mississippi delta (89.2$^\circ$W, 29$^\circ$N) often has an agglomeration of highly-attractive cLCSs; during spring and summer, the shelf just to the east (85-89$^\circ$W,$\sim$30$^\circ$N) has many cLCSs with high c$\rho$ values, relative to other shelves (see also Fig. \ref{fig:fig01b} in the Supplemental Information). This agrees well with transport and beaching of DwH oil, during April-August of 2010\cite{LeHenaff2012,Weisberg2017}. It also agrees with several studies using  multi-year trajectory simulations to determine the likely outcome for a spill originating at the Macondo well under spring and summer conditions: The vicinity of the Mississippi delta is most-at-risk, followed by the shelf and coast to the east up to about 85$^\circ$W; the LC is also found to attract trajectories in these multi-year simulations, although oil from the DwH did not reach it \cite{Barker2011,Ji2011,Tulloch2011}. These studies use velocity fields from years 1992-2008, 1993-1998 and 1992-2007, respectively, to compute probability of impact based on a point-source oil spill, and forward-in-time integrations of the instantaneous velocity. Thus, while the model we use requires due caution when interpreting coastal circulation,  our results seem to confirm these studies, and the first part of a conclusion in Weisberg et al.\cite{Weisberg2017} (who  studied DwH oil beaching using different models and observations): ``In essence it is found that the circulation gets the oil to the vicinity of the beach, whereas the waves, via Stokes drift, are responsible for the actual beaching of oil.''.\\
\noindent
Our analysis identifies that the interior of the three wide shelves of the GoM --the West Florida, La-Tex and Yucatan shelves-- are isolated throughout the year; the 50-m isobath being a good indicator for the transport barrier. Shallower than this isobath, c$\rho$ is for the most part negligible, implying low stirring activity, and that water parcels within the shelves are unlikely to have originated from outside. This is consistent with what is expected from wide shelves \cite{Brink2016}. However, some cLCSs with strong c$\rho$ values can be seen sporadically within some of these shelves, e.g. in the Yucatan shelf near (91$^\circ$W, 21$^\circ$N), specially during the summer. Also, highly-attractive cLCSs may persist near the coastline. \\
\noindent
The isolation of the West Florida shelf  has been documented  from observations and numerical models \cite{Yang-etal-99,Olascoaga-etal-06,Olascoaga-NPG-10}, while observational and numerical studies have also noted that the vicinity of the 50-m isobath separates distinct kinematical and dynamical regimes within that shelf \cite{Sturges2001,Li1999a,Li1999b}. Thus, our results seem a confirmation of the underlying dynamics described in these studies.
\noindent
The cLCSs reported by Gough et al.\cite{Goughetal2017} accurately identify transport barriers and isolation for the La-Tex shelf, as evaluated through comparisons with synthetic drifters (advected  by their instantaneous velocity) and historical satellite-tracked drifters. Thus providing an independent confirmation for isolation of the La-Tex shelf. Isolation for the Yucatan shelf will be illustrated below, with observations from the Ixtoc oil spill. \\
Highly-attractive cLCSs persist along the western coastline of the GoM, this pattern is consistent with the coastal vulnerability (attraction of synthetic drifters) found by Thyng \& Hetland\cite{ThyngHetland2017}, from about 24 to 29$^\circ$N   (cf. their Figs. 3 and 8). And as we will see, it is also consistent with oil beaching from the Ixtoc oil spill.

\subsection*{Direct comparison with observations.}
\label{direct}
\subsubsection*{The ``tiger tail''.}
\begin{figure} 
\begin{center}
  \includegraphics[scale=1.70]{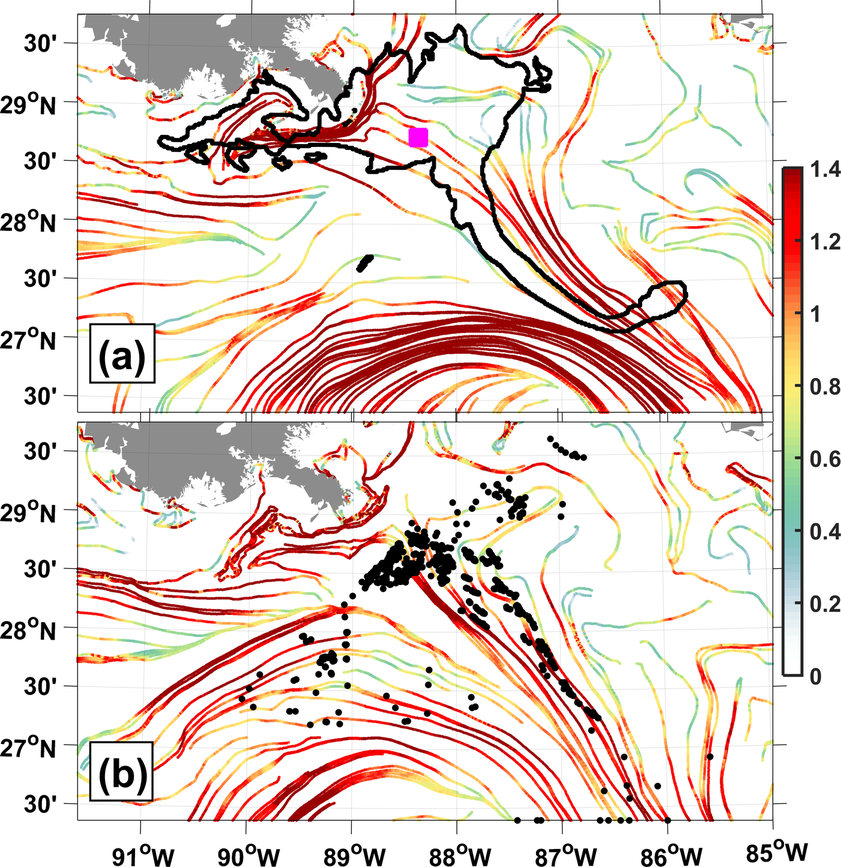}%
  \caption{ 
   Climatological LCSs colored according to their climatological attraction strength  $\ln \mathrm{c}\rho$  for (a) May and (b) July in the northern GoM. Superimposed in (a) is the oil outline (black) from the Deepwater Horizon spill as seen on May 17, 2010; the Macondo well location (magenta square) is also shown.  In (b), black dots are the daily positions of GLAD drifters from July 29 to August 2, 2012.   }
  \label{fig:fig02}
\end{center}
\end{figure}

\noindent
During the DwH spill in May 2010, a current resembling a localized jet was responsible for a significant redistribution of the sea-surface oil slick.  The resulting prominent filament became known as a ``tiger tail'' \cite{Olascoaga2012}.  The tiger tail stretched along the direction indicated by the cLCSs for May (Fig. \ref{fig:fig02}a). We also note that between 28.5-29.25$^\circ$N and 89.5-91.5$^\circ$W, the oil outline closely conforms to the cLCSs just south and west of the Mississippi delta (Fig. \ref{fig:fig02}a). \\ 
For the most part of July 2012, a filament similar to the tiger tail was observed with GLAD drifters, sea-surface temperature and chlorophyll\cite{Olascoaga-etal-13}; July's cLCSs accurately show the direction of stretching (Fig. \ref{fig:fig02}b).

\subsubsection*{Ixtoc oil spill.}
\begin{figure}
\begin{center}
\includegraphics[width=\textwidth]{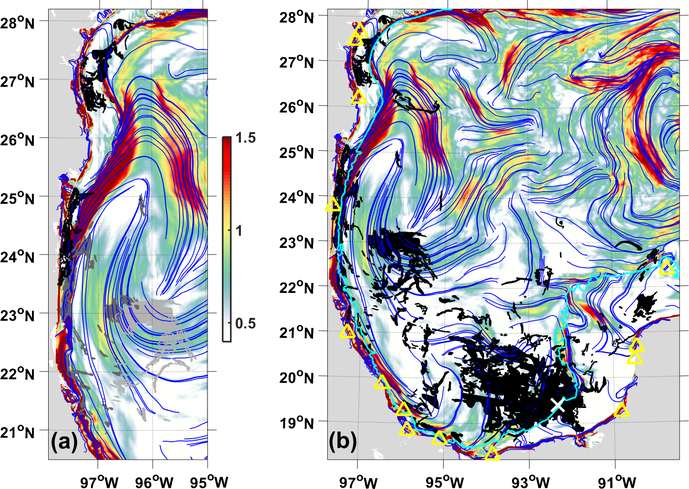}%
  \caption{Climatological LCSs (blue curves) on climatological attraction strength  $(\ln \mathrm{c}\rho$)  for August in the western GoM. (a) Superimposed in the left panel is oil from the \emph{Ixtoc} spill as observed (Landsat/MSS satellite sensor) on August 1 (light gray), August 2 (gray) and August 21 (black), 1979. (b) Superimposed in the right panel is oil (black) from the \emph{Ixtoc} spill as observed (Landsat/MSS and CZCS satellite sensors) through the almost 10-month duration of the blowout. The location of the blowout (92.33$^\circ$W,19.41$^\circ$N) is marked with a white cross and the 50-m isobath is plotted in cyan. Confirmed beachings reported by Sun et al.\cite{Sun2015} are marked with yellow triangles.}
  \label{fig:fig03}
  \end{center}
\end{figure}

\noindent
Similar to the DwH accident, the \emph{Ixtoc} rig off the coast of Mexico exploded in June 1979 (the blowout location is marked in Fig. \ref{fig:fig03}b).  The well was capped in March 1980, spilling the second largest accidental oil release after the DwH incident. Oil trajectories had two salient characteristics in that event: firstly the Yucatan shelf just east of the well remained relatively isolated; and secondly, the oil with the longest trajectories moved north and west, impacting the western GoM coast \cite{Sun2015}. We find these patterns in the cLCS fields for trajectories originating in the vicinity of the accident. And again, this is consistent with the northward advection along the western margin that was reported by Zavala-Sanson et al. \cite{Zavala-Sanson-etal-17}, as a dominant dispersion scenario for point sources near the Ixtoc blowout. \\
\noindent
Oil moving northwestward was documented in a two-day sequence of satellite images starting on August 1, 1979 (Fig. \ref{fig:fig03}a). On the first day, oil is positioned over August's cLCSs that direct westward and then northward at about 95-97$^\circ$W, 22-23$^\circ$N. The satellite image next day (August 2) shows that this oil moved westward and northward following these cLCSs;  note that not much filamentation is observed, possibly due to cloud coverage. In this second-day image, oil can also be seen closer to the coast aligned with about 300-km of highly-attractive cLCSs,  following the 50-m isobath (the isobath is shown in Fig.\ref{fig:fig03}b). However, it seems unlikely that oil stretching along cLCSs over the the 50-m isobath on August 2,  made it from the oil observed on the previous day at about 23$^\circ$N  -- it would have required a velocity of one to two meters per second towards the coast. This suggests that on August 1, oil was already closer to the coast, but covered by clouds. The next available image with Ixtoc oil is on August 29, 1979. In this image, oil can be seen even closer to the coast at about 23.5-25$^\circ$N and 26.5-28$^\circ$N, near highly-attractive coastal cLCSs (which can also be seen in Fig. \ref{fig:fig01a}). These coastal cLCSs with with c$\rho$ maxima, are consistent with the coastal vulnerability (attraction of synthetic drifters towards the coastline) reported by Thyng \& Hetland \cite{ThyngHetland2017}, as mentioned above.\\
\noindent
When all the available satellite observations from the Ixtoc spill are plotted together, the 50-m isobath of the Yucatan shelf emerges as an effective barrier, although with some exceptions. The spill originated just off the 50-m isobath less than 100km west of the Yucatan shelf. However, only a small amount of oil --relative to the amount to the north, west and northwest of the blowout-- moved eastward onto the Yucatan shelf (Fig. \ref{fig:fig03}b). Oil  moved southeast from the blowout towards highly-attractive coastal cLCSs near (91.8$^\circ$W,18.7$^\circ$N), along weakly-attracting cLCSs -- this constitutes the most significant example of cross-shelf transport in these data.   With one exception, all confirmed beachings along the Mexican and U.S. coasts happened where near-coast cLCSs have c$\rho$ maxima (Fig. \ref{fig:fig03}b; see also Fig. \ref{fig:fig01a}).

\subsection*{Time-variability of the transport patterns.}
\noindent
The temporal variability of the cLCSs can be assessed by superimposing the LCSs from the twelve dynamical systems over which the CG-tensor averaging takes place.  In the Supporting Information (Appendix B) we show quantitatively that the cLCSs are effectively equivalent to the superposition of LCSs from the dynamical systems in the averaging period. The LCSs superposition confirms the transport patterns described above.\\
The similarity between a month's cLCSs and the superposition of LCSs from 7-day flow maps spanning that month, suggests that the climatological velocity does not have much temporal variability over 30-day periods. However, our method cannot be simplified further by first monthly-averaging the climatological velocity and then computing streamlines, as we show in Appendix C of the Supporting Information. \\
\noindent
It is when a suitable low-pass filter has been applied to the velocity field that recurrent, coherent pathways may emerge.  When using the instantaneous velocity, the LCSs from the different dynamical systems have comparable stretching strengths, but orientation is more varied, and the cLCSs resemble a disorganized collection of LCSs.\\
One cLCS map per month is enough to anticipate a range of known circulation patterns. Several of the cLCS maps are qualitatively similar from month to month. The spatial location and strength of the LCS from different 7-day dynamical systems spanning a month are very similar to the corresponding cLCSs. These characteristics confirm the quasi-steady nature of our results. Indeed, Gough et al.\cite{Goughetal2017} found good agreement between simulated and observed transport patterns and cLCSs computed from yearly-averaging 7-day CG tensors.\\
\noindent

\section*{Concluding remarks.}
\noindent
We have presented several independent confirmations that the cLCSs extract kinematics from multi-year time-series, synthesizing important information. cLCSs strongly constrain the ensemble-mean transport sustained by the instantaneous circulation, thus linking the climatological and instantaneous velocity fields. To the best of our knowledge, this is the first time that a velocity climatology has been shown to preserve transport information from the instantaneous velocity. Our results agree with data beyond the time period spanned by our velocity climatology, revealing that persistent, or recurrent, circulation has been found.  Additionally, known prominent transport in the GoM are accurately depicted, including some of the highest-profile circulation patterns observed in recent history. These characteristics suggest the utility of our Lagrangian transport climatology for a variety of applications.  As examples, those planning for environmental-pollution prevention and response may use the monthly cLCS maps to identify regions at high risk of being visited by contaminants, depending on the location of a pollution source.  Regions that are isolated or stagnated are unlikely to be impacted if the pollution source lies outside, but will be heavily impacted if the source is within. Likely transport patterns can also be deduced, e.g. the tiger-tail filaments in the vicinity of the DwH. Our method generalizes the commonly used approach of running multiple simulations to compute the probability of contact given a point source. This is because cLCS maps classify regions' risk of contact given a generic oil spill: Our results do not require, and therefore do not depend on, knowing the source's location \emph{a priori}. Also, our method provides more information relative to the probability-of-contact approach, because it clearly depicts likely pathways. These types  of risk maps provide valuable information complementing oil-spill forecasts  (e.g. Barker\cite{Barker2011}). We emphasize that, by definition, a climatology discards weather patterns; our results are therefore a complement and not a substitute for instantaneous-velocity simulations.\\ 
\noindent
 Other applications for the cLCS method arise from recognizing the persistent or recurrent nature of kinematics as seen through cLCSs, say, e.g., for determining ideal navigation routes in a climatological sense, or possibly by gaining insight into the nature of the dynamics responsible for the low-frequency kinematics depicted by cLCSs.  \\
\noindent
Our work shows that it is possible to find quasi-steady, general patterns that describe important aspects of the inherently time-dependent, chaotic problem of oceanic Lagrangian transport -- this appraises new applications as practical.
 
\section*{Acknowledgments.}
\noindent
The work of RD was supported by the National Energy Technology Laboratory's ongoing research under the Offshore Field Work Proposal DOE NETL FY14-17 under the RES contract DE-FE0004000.  The work of FJBV and MJO was supported by the Gulf of Mexico Research Initiative as part of the Consortium for Advanced Research on Transport of Hydrocarbon in the Environment (CARTHE) and SENER--CONACyT grant 201441 as part of the Consorcio de Investigaci\'on del Golfo de M\'exico (CIGoM).
RD thanks Prof. George Haller for helpful conversations, and Prof. Roger M. Samelson for insightful comments on earlier versions of this manuscript. We thank Shaojie Sun for providing the Ixtoc oil-spill satellite data. The HYCOM$+$NCODA $1/25^\circ$ GoM Reanalysis was funded by the U.S.\ Navy and the Modeling and Simulation Coordination Office and is available at hycom.org. Computer time was made available by the DoD High Performance Computing Modernization Program. The GLAD drifter trajectory dataset is publicly available through the Gulf of Mexico Research Initiative Information \& Data Cooperative (GRIIDC) at https://data.\allowbreak
gulfresearchinitiative.\allowbreak org  (DOI:10.7266, N7VD6WC8). NOAA overflight information is available at http://\allowbreak
www.ssd.noaa.gov/\allowbreak PS/MPS/\allowbreak deepwater.html. jLab was used for data pre-processing and plotting (http://www.jmlilly.net/jmlsoft.html).

\section*{Author contributions.}
The authors proposed this study independently, and coincidentally ended 
working on it together. Velocity data was acquired and pre-processed by RD. Computations were done by RD using  code written by RD, FJBV, MJO, and Florian Huhn; except for the ensemble-mean subsection of the results, computed by FJBV.  All authors contributed towards developing the techniques and interpreting results. RD wrote the paper with contributions by FJBV and MJO. 

\section*{Additional Information.}

{\bf Supplementary information} accompanies this paper.\\
\noindent
{\bf Competing financial interests:} The authors declare that they have no competing financial interests.

\noindent
{\bf Disclaimer.} This project was funded by the Department of Energy, National Energy Technology Laboratory, an agency of the United States Government, through a support contract with AECOM. Neither the United States Government nor any agency thereof, nor any of their employees, nor AECOM, nor any of their employees, makes any warranty, expressed or implied, or assumes any legal liability or responsibility for the accuracy, completeness, or usefulness of any information, apparatus, product, or process disclosed, or represents that its use would not infringe privately owned rights. Reference herein to any specific commercial product, process, or service by trade name, trademark, manufacturer, or otherwise, does not necessarily constitute or imply its endorsement, recommendation, or favoring by the United States Government or any agency thereof. The views and opinions of authors expressed herein do not necessarily state or reflect those of the United States Government or any agency thereof. \\

\newpage
\section*{Appendix Contents}

\begin{itemize}

\item Appendix A. Divergence of the climatological velocity.

\item Appendix B. A quantitative comparison between cLCSs and LCSs.   
  
\item Appendix C. Comparing cLCSs with the streamlines for the monthly-averaged climatological velocity.

\end{itemize}

\appendix

\section{Divergence of the climatological velocity.}
\begin{figure}[h]
\centering
  \includegraphics[scale=1.5]{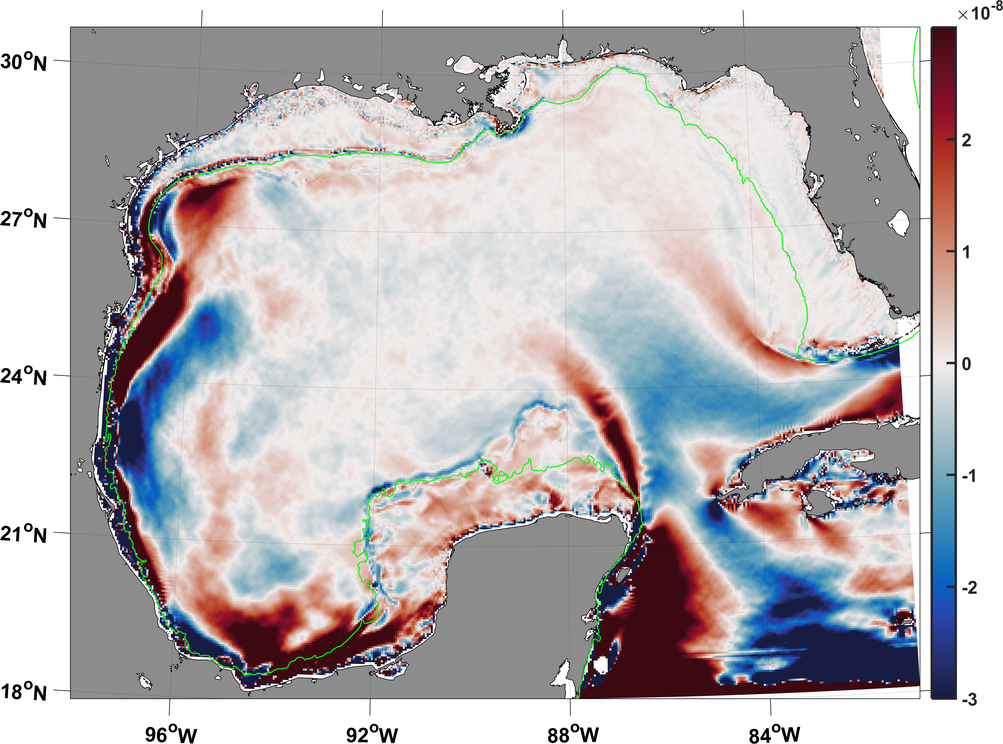}
\caption{Yearly average of the Eulerian divergence (day$^{-1}$) computed from the daily values in our climatological velocity (described in the paper). The 50-m isobath is shown in green. Saturated colors are an order of magnitude bigger than the shown colorscale.}
\label{fig:eul_div}
\end{figure}

\noindent
In this section we investigate if the cLCSs' attraction is due to confluence (i.e. divergence-free attraction) or convergence (attraction with negative divergence). We use the notation described in the paper.
\noindent
Let $\delta\left(\mathbf{x},t\right) \coloneqq \nabla \cdot \mathbf{v}\left(\mathbf{x},t\right)$ be the Eulerian divergence of our two-dimensional climatological velocity field (described in the paper). Daily values of the Eulerian divergence from our climatology have several persistent features that are aptly captured in the yearly mean: These characteristic features include strong positive divergence with negative divergence next to it along the western boundary of the GoM, positive divergence between 92-96 $^\circ$W at about 19$^\circ$N, and the LC is characterized by positive divergence to both sides of negative divergence  (Fig. \ref{fig:eul_div}). \\
\noindent
We are interested in the divergence along the paths we used for our cLCSs computations, to understand if the back-in-time repulsion we compute can be associated with an increase in sea-surface area due to positive divergence. In forward time this would imply that the attraction we report is associated with an area decrease due to negative divergence. 
\noindent
We can use the equation for the change of a material area $\mathrm{d} A(t) / \mathrm{d} t = \delta(t) A(t)$ which has the solution 
\begin{equation}
A\left(t\right)=A_0 \exp \left(\int_{t_0}^t \delta \left(\mathbf{F}^{t'}_{t_0}\left(\mathbf{x}_0\right),t' \right) \mathrm{d} t' \right)
\label{area}
\end{equation}

\noindent
Define the parameter 
\begin{equation} \alpha \coloneqq A(t)/A_0 \label{alpha}\end{equation}
(These equations are often treated in textbooks of classical mechanics under Liouville's theorem and subsequent lemmas.)
We compute $\alpha$ by evaluating the exponential in \eqref{area} using the same flow maps $\mathbf{F}^{t}_{t_0}\left(\mathbf{x}_0\right)$ that we used for the computation of cLCSs (described in the paper). Note that the fractional change of area is the Jacobian determinant of the transformation $\mathbf{F}^t_{t_0}$, i.e. $\alpha=\det \left(\mathrm{D}\mathbf{F}^t_{t_0}\right)$. When $\alpha\approx 1$, shrinking or expanding of areas is negligible over the integration period and attraction (in forward time) would be due to confluence. \\ 
\begin{figure}[h]
\centering
  \includegraphics[width=\textwidth]{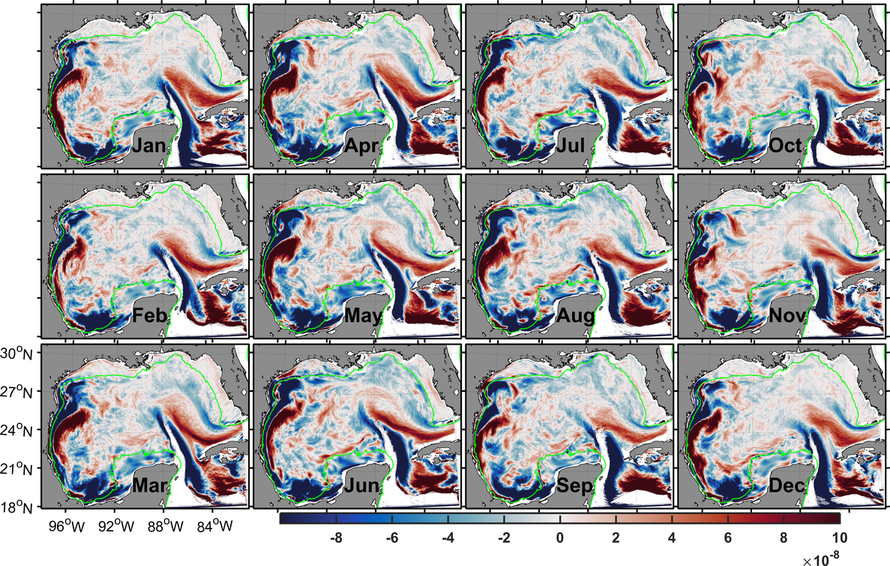}
\caption{Monthly-averages of the parameter $\alpha$ defined in \eqref{alpha}, and using the same flow maps as in the computation of cLCSs. The colorscale represents the signed distance from one, e.g., a value of $-2\times10^{-8}$ represents $1-2\times10^{-8}=0.99999998$. The monthly averages are computed by averaging the values of $\alpha$ from each of the twelve dynamical systems in a month (as described in the paper). $\alpha$ is plotted as a function of the trajectories' initial position $\mathbf{x}_0$  used to compute the exponential in \eqref{area}. The 50-m isobath is shown in green. Saturated colors are an order of magnitude bigger than the shown colorscale.}
\label{fig:lag_div}
\end{figure}

\begin{figure}[h]
\centering
  \includegraphics[scale=0.45]{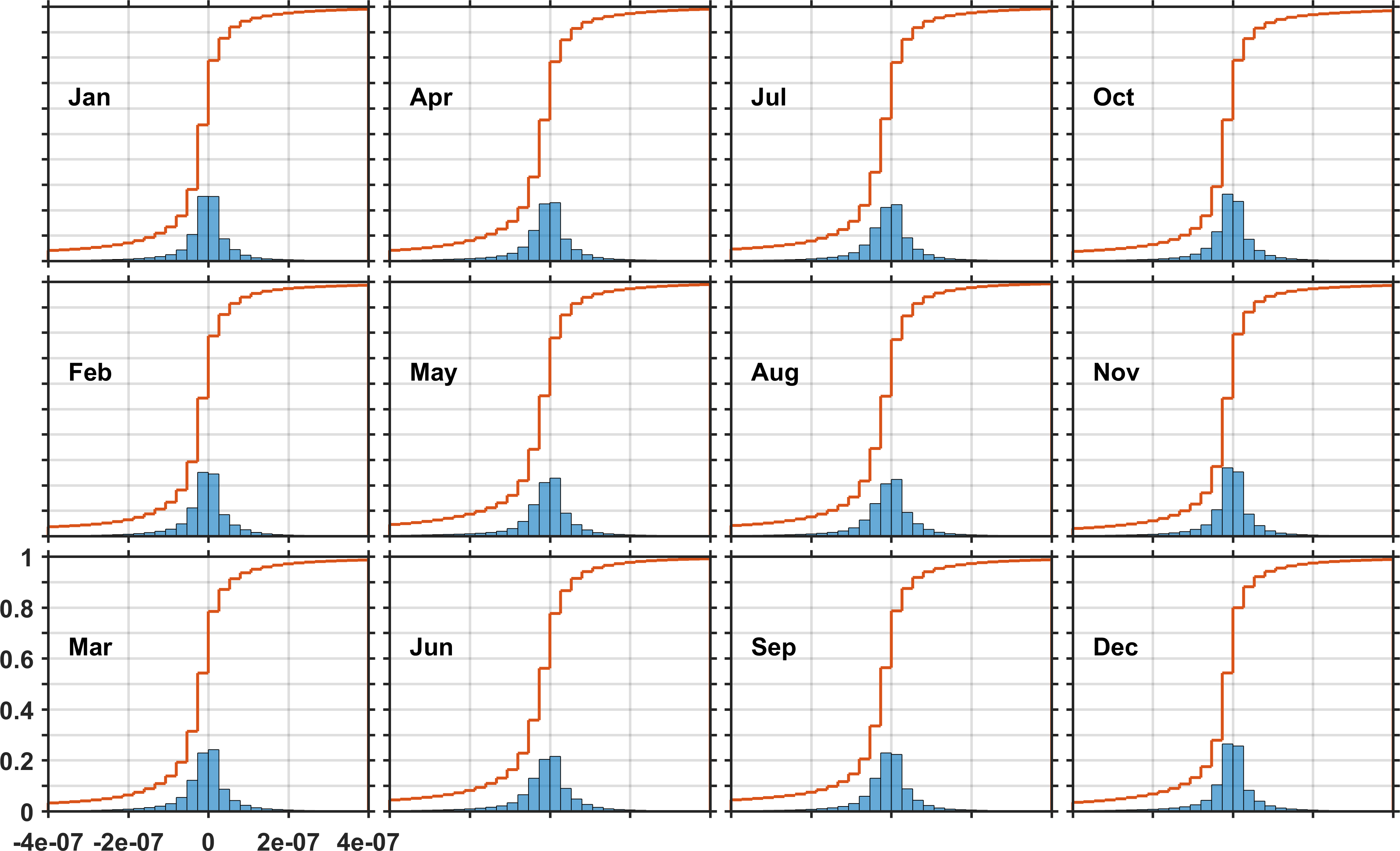}
\caption{Probability distribution (blue bars, abscissa) and cumulative probability distribution (orange lines, abscissa) for all the  values of $\alpha$ (defined in \eqref{alpha}) in each month, as a function of the signed distance from one (ordinate; e.g., a value of $-1\times10^{-7}$ represents $1-1\times10^{-7}=0.9999999$). Values for each month correspond to all the values of $\alpha$ from the twelve dynamical systems spanning that month that were used for the cLCS computation described in the paper.}
\label{fig:lag_div_hist}
\end{figure}

\noindent 
We compute monthly-mean fields by averaging the twelve  $\alpha$ fields, obtained from the flow maps of the twelve dynamical systems we used to characterize each month (Fig. \ref{fig:lag_div}).
\noindent
A closer look at the values of $\alpha$ for each month is presented in table 1. Even the smallest (0.9995) and biggest values (1.00005) of $\alpha$ throughout our climatology, result in negligible changes of area. The absolute maximum reduction of area as it was advected with the flow map corresponds to 0.05\% of the original area; the absolute maximum increase is 0.005\% of the original area. Furthermore, most values of $\alpha$ are much closer to one than these extreme values, as can be seen in the probability distributions (Fig. \ref{fig:lag_div_hist}). We note that distributions are skewed towards negative values (Table \ref{table:stats} and Fig. \ref{fig:lag_div_hist}).  

\noindent
In these back-in-time integrations, negative values work to impede back-in-time repulsion and therefore forward-in-time attraction. Thus, we conclude that the effect of divergence is negligible, and that out of the negligible effect, the most significant part acts to counter the attraction we report. \\

\begin{center} 
\vspace{0.08 in}
\begin{tabular}{cccccccc}
Month & Mean & StdDev & AbsMin & 1$^\mathrm{st}$Quartile & Median & 3$^\mathrm{rd}$Quartile & AbsMax \\             
 \hline
 \hline
Jan & -3e-07 & 5e-06 & -3e-04 & -3e-08 & -3e-09 & 2e-08 & 4e-06 \\        
Feb & -3e-07 & 5e-06 & -3e-04 & -3e-08 & -3e-09 & 2e-08 & 5e-06 \\        
Mar & -4e-07 & 7e-06 & -3e-04 & -4e-08 & -4e-09 & 2e-08 & 5e-06 \\        
Apr & -4e-07 & 7e-06 & -5e-04 & -4e-08 & -5e-09 & 2e-08 & 1e-05 \\        
May & -4e-07 & 7e-06 & -5e-04 & -4e-08 & -6e-09 & 2e-08 & 6e-06 \\        
Jun & -5e-07 & 8e-06 & -5e-04 & -5e-08 & -7e-09 & 2e-08 & 4e-06 \\        
Jul & -4e-07 & 6e-06 & -4e-04 & -5e-08 & -7e-09 & 2e-08 & 3e-06 \\        
Aug & -4e-07 & 7e-06 & -4e-04 & -4e-08 & -6e-09 & 2e-08 & 7e-06 \\        
Sep & -4e-07 & 6e-06 & -4e-04 & -4e-08 & -6e-09 & 2e-08 & 5e-05 \\        
Oct & -3e-07 & 5e-06 & -2e-04 & -3e-08 & -5e-09 & 2e-08 & 2e-05 \\        
Nov & -2e-07 & 4e-06 & -3e-04 & -3e-08 & -3e-09 & 2e-08 & 5e-05 \\        
Dec & -2e-07 & 4e-06 & -3e-04 & -3e-08 & -3e-09 & 2e-08 & 6e-06    
\end{tabular}                                                                          
\captionof{table}{Statistics for all the values of $\alpha$ spanning the twelve dynamical systems in each month; values are presented as the signed distance from one, e.g. -3e-07 represents a value of 1-3e-07=0.9999997. \label{table:stats}}                                                                                                                                                                                       
\end{center}
\vspace{0.2 in}
\section{A quantitative comparison between cLCSs and LCSs.}
\begin{figure}[h]
\centering
  \includegraphics[width=\textwidth]{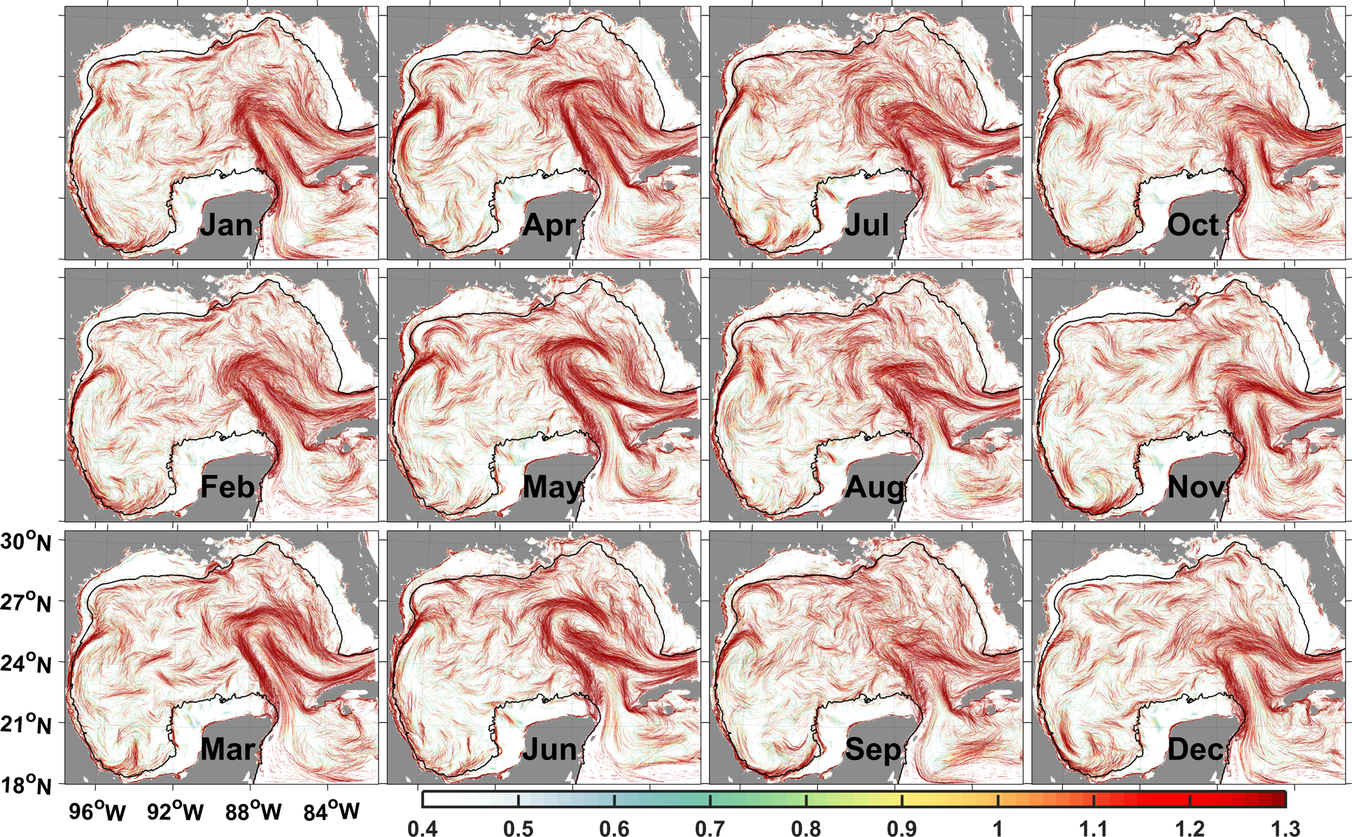}%
  \caption{Superposition of the LCSs from the 12 dynamical systems used to compute the monthly-averaged Cauchy-Green tensor from where climatological LCS are computed (shown in Fig. 2 of the paper). Colors represents the LCSs' attraction strength $\ln \rho$.  The 50-m isobath is indicated in black.  }
  \label{fig:fig01b}
\end{figure}
\begin{figure}[h]
\centering
  \includegraphics[scale=0.75]{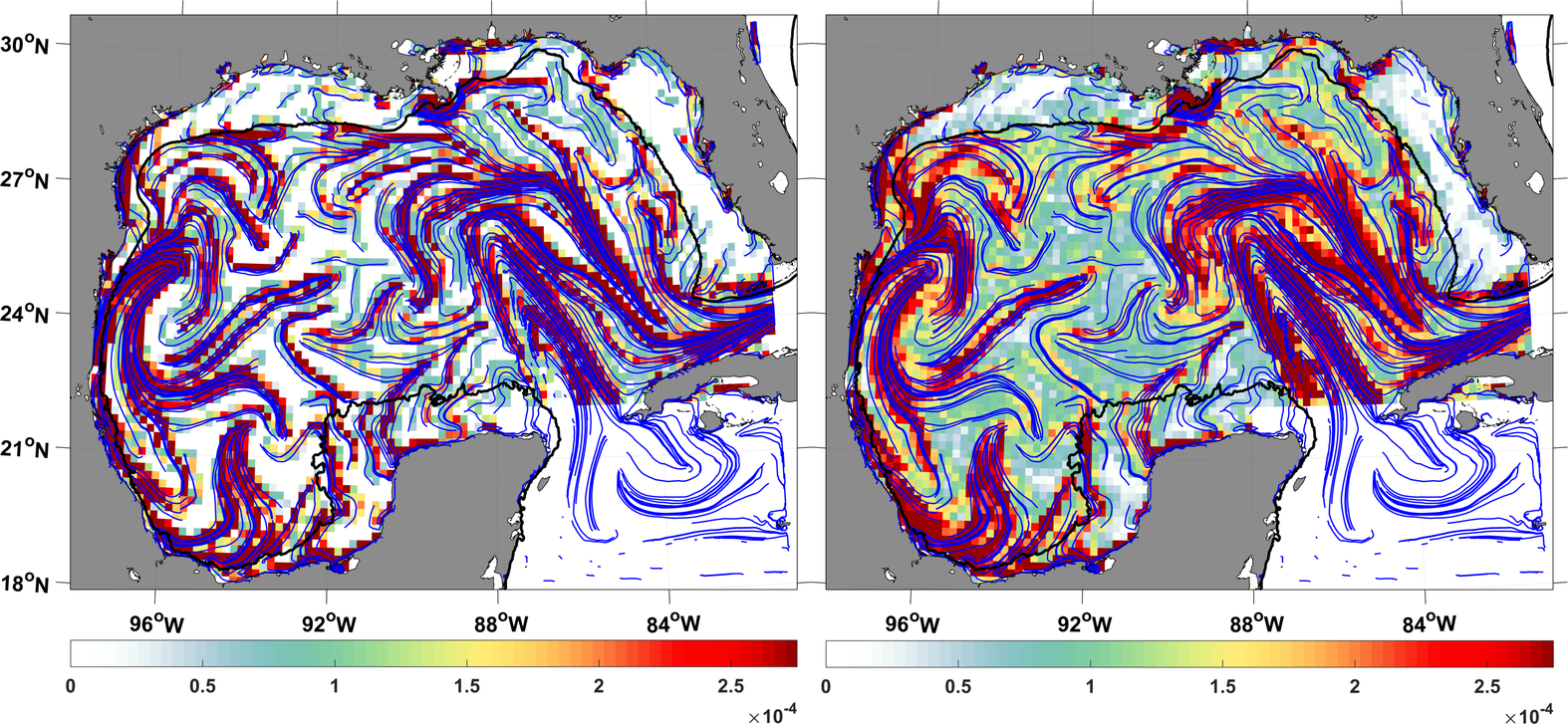}%
  \caption{Both panels show the climatological Lagrangian Coherent Structures (cLCSs; blue lines) for April. In colors, the left panel shows the probability distribution (on a $0.175\times0.175$ degree grid) of April's cLCSs $(x,y)$ positions. The right panel is the probability distribution (same grid) of the $(x,y)$ positions of all the LCSs from each of the CG tensors that were averaged to get April's cLCSs. The 50-m isobath is shown in black. Plots for all months are qualitatively similar, therefore, only  the comparison for the month with the smallest correlation is shown; correlations for all months are presented in table 2.}
  \label{fig:lcs}
\end{figure}

cLCSs are the squeezelines of a monthly-mean Cauchy-Green (CG) tensor; the averaging is described in the paper for which this is a supplement. Here we show that the LCSs from the twelve CG tensors over which we average, often persist in similar positions, and that the corresponding cLCSs are similarly located.  
Our approach is to compare the cLCSs to the superposition of LCSs from all of the CG tensors over which we average when computing the monthly-mean CG tensor. A visual inspection between the superposition of all LCSs (Fig. \ref{fig:fig01b}) readily suggests a striking  similarity to the corresponding cLCSs (compare to Fig. 2 of the paper).
\noindent
For a quantitative analysis, LCSs and cLCSs are first interpolated to equidistant points along the curve, making them directly comparable. We then use a $0.175 \times 0.175$ degree grid to compute the probability that any given cell contains cLCS $(x,y)$ points, and compare that to the probability that the same cell has LCS $(x,y)$ points from any of the dynamical systems in the averaging period (Fig. \ref{fig:lcs}).  A robust least squares regression shows a significantly-correlated linear relation for the cell-wise comparison of the two probability distributions. The details for each month's linear regression are shown in table 2.\begin{center}
\begin{tabular}{ccc} 
Month & Correlation & Slope \\ 
\hline 
\hline
Jan & 0.74 & 0.55 \\ 
Feb & 0.75 & 0.47 \\ 
Mar & 0.77 & 0.59 \\ 
Apr & 0.64 & 0.49 \\ 
May & 0.74 & 0.53 \\ 
Jun & 0.71 & 0.52 \\ 
Jul & 0.73 & 0.46 \\ 
Aug & 0.7 & 0.46 \\ 
Sep & 0.76 & 0.47\\ 
Oct & 0.76 & 0.48 \\ 
Nov & 0.72 & 0.5 \\ 
Dec & 0.76 & 0.51 
\end{tabular} 
\label{table:1}
\captionof{table}{Correlation between the cLCSs and LCSs probabilities for each month,  values above 0.05 are significant with 95\% confidence. The slope of the robust least-squares linear fit is also shown. All y-intercepts are between 5 and 7 $\times 10^{-5}$. } 
\end{center}

\noindent
Effectively, LCSs from the climatological velocity identify quasi-steady transport patterns over a month time period, and computing cLCSs is an efficient way of extracting these patterns.


\section{Comparing cLCSs with the streamlines for the climatological velocity's monthly mean.}
\begin{figure}[h]
\centering
  \includegraphics[scale=0.30]{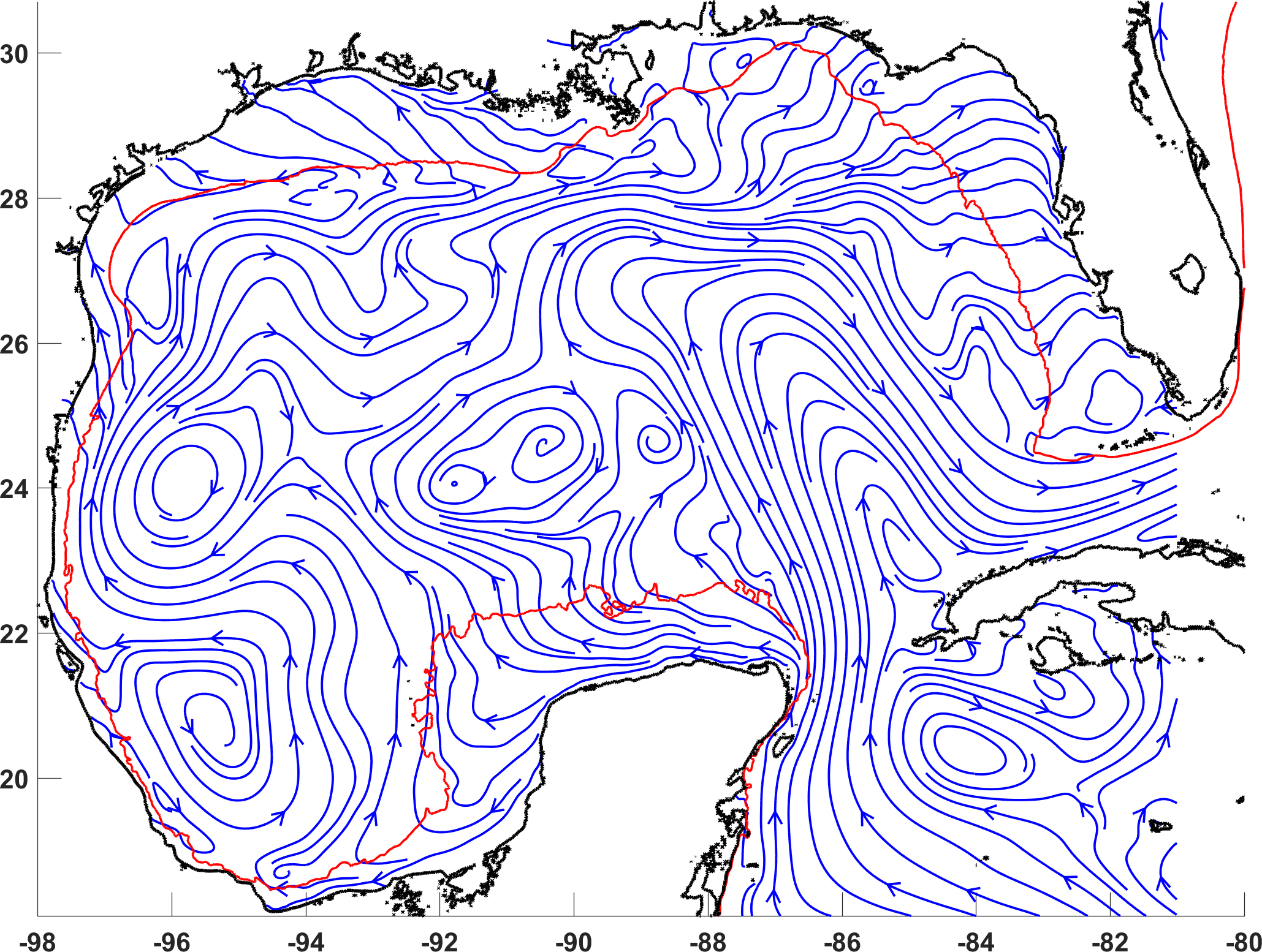}%
  \caption{Streamlines (blue lines) for April computed from the climatological velocity that was used to compute April's cLCSs after monthly averaging; the 50-m isobath is shown in red.  Notice the cross-shore flow throughout the West Florida and other shelves; compare to the cLCSs for April shown in Fig. \ref{fig:fig01a} of the paper, and also the superposition of LCSs shown in Fig. \ref{fig:fig01b} of this supplement. }
  \label{fig:streamlines}%
\end{figure}

\noindent
In the previous section we showed that the climatological velocity does not have much variability within a month timescale in the sense that the LCSs from the different dynamical systems spanning any month are very similar to the corresponding cLCSs. The purpose of this section is to show that our analysis cannot be simplified further by computing streamlines of the monthly-averaged climatological velocity, instead of the cLCSs that we compute.  For example, April's monthly-averaged climatological-velocity streamlines show cross-shelf transport in Florida's western and northwestern shelves (Fig.\ \ref{fig:streamlines}). In contrast,  April's cLCSs and LCSs show an isolated  West Florida Shelf (cf.\ Fig. \ref{fig:fig01a} of the paper and Fig. \ref{fig:fig01b} of this supplement), in agreement with previous observational and numerical studies (described in  the paper). The Yucatan and La-Tex shelves also show unrealistic cross-shore transport. Other months also show spurious transport patterns.  Furthermore, the streamlines do not identify regions of isolation or stagnation, which are accurately identified by regional minima of c$\rho$.

\begin{thebibliography}{10}
\expandafter\ifx\csname url\endcsname\relax
  \def\url#1{\texttt{#1}}\fi
\expandafter\ifx\csname urlprefix\endcsname\relax\def\urlprefix{URL }\fi
\expandafter\ifx\csname doiprefix\endcsname\relax\def\doiprefix{DOI }\fi
\providecommand{\bibinfo}[2]{#2}
\providecommand{\eprint}[2][]{\url{#2}}

\bibitem{LaCasce2008}
\bibinfo{author} {LaCasce, J. H.},
\newblock \bibinfo{journal}{\bibinfo{title}{Statistics from Lagrangian observations}}.
\newblock {\emph{\JournalTitle{Progress in Oceanography}}}
\textbf{\bibinfo{volume}{77}}
\bibinfo{pages}{1--29}
(\bibinfo{year}{2008}).
\newblock \doiprefix 10.1016/j.pocean.2008.02.002 

\bibitem{Haller-Yuan-00}
\bibinfo{author}{Haller, G.} \& \bibinfo{author}{Yuan, G.}
\newblock \bibinfo{journal}{\bibinfo{title}{Lagrangian coherent structures and
  mixing in two-dimensional turbulence}}.
\newblock {\emph{\JournalTitle{Physica D}}} \textbf{\bibinfo{volume}{147}},
  \bibinfo{pages}{352--370} (\bibinfo{year}{2000}).

\bibitem{Samelson2013}
\bibinfo{author}{Samelson, R.~M.}
\newblock \bibinfo{journal}{\bibinfo{title}{{Lagrangian motion, coherent
  structures, and lines of persistent material strain.}}}
\newblock {\emph{\JournalTitle{Annual review of marine science}}}
  \textbf{\bibinfo{volume}{5}}, \bibinfo{pages}{137--163}
  (\bibinfo{year}{2013}).
\newblock \urlprefix\url{http://www.ncbi.nlm.nih.gov/pubmed/22809180}.
\newblock \doiprefix 10.1146/annurev-marine-120710-100819.

\bibitem{Haller2015}
\bibinfo{author}{Haller, G.}
\newblock \bibinfo{journal}{\bibinfo{title}{{Lagrangian Coherent Structures}}}.
\newblock {\emph{\JournalTitle{Annual Review of Fluid Mechanics}}}
  \textbf{\bibinfo{volume}{47}}, \bibinfo{pages}{140906185740003}
  (\bibinfo{year}{2015}).
\newblock \urlprefix\url{www.annualreviews.org
  http://www.annualreviews.org/doi/abs/10.1146/annurev-fluid-010313-141322}.
\newblock \doiprefix 10.1146/annurev-fluid-010313-141322.

\bibitem{Haller-11}
\bibinfo{author}{Haller, G.}
\newblock \bibinfo{journal}{\bibinfo{title}{A variational theory of hyperbolic
  {Lagrangian Coherent Structures}}}.
\newblock {\emph{\JournalTitle{Physica D}}} \textbf{\bibinfo{volume}{240}},
  \bibinfo{pages}{574--598} (\bibinfo{year}{2011}).
\newblock \doiprefix 10.1016/j.physd.2010.11.010.

\bibitem{Olascoaga-etal-13}
\bibinfo{author}{Olascoaga, M.~J.} \emph{et~al.}
\newblock \bibinfo{journal}{\bibinfo{title}{{Drifter motion in the Gulf of
  Mexico constrained by altimetric Lagrangian Coherent Structures}}}.
\newblock {\emph{\JournalTitle{Geophys. Res. Lett.}}}
  \textbf{\bibinfo{volume}{40}}, \bibinfo{pages}{6171--6175}
  (\bibinfo{year}{2013}).
\newblock \doiprefix 10.1002/2013GL058624.

\bibitem{Haller-Beron-12}
\bibinfo{author}{Haller, G.} \& \bibinfo{author}{Beron-Vera, F.~J.}
\newblock \bibinfo{journal}{\bibinfo{title}{Geodesic theory of transport
  barriers in two-dimensional flows}}.
\newblock {\emph{\JournalTitle{Physica D}}} \textbf{\bibinfo{volume}{241}},
  \bibinfo{pages}{1680--1702} (\bibinfo{year}{2012}).
\newblock \doiprefix 10.1016/j.physd.2012.06.012.

\bibitem{Farazmand-etal-14}
\bibinfo{author}{Farazmand, M.}, \bibinfo{author}{Blazevski, D.} \&
  \bibinfo{author}{Haller, G.}
\newblock \bibinfo{journal}{\bibinfo{title}{Shearless transport barriers in
  unsteady two-dimensional flows and maps}}.
\newblock {\emph{\JournalTitle{Physica D}}} \textbf{\bibinfo{volume}{278-279}},
  \bibinfo{pages}{44--57} (\bibinfo{year}{2014}).

\bibitem{Beron-etal-15}
\bibinfo{author}{Beron-Vera, F.~J.} \emph{et~al.}
\newblock \bibinfo{journal}{\bibinfo{title}{{Dissipative inertial transport
  patterns near coherent Lagrangian eddies in the ocean}}}.
\newblock {\emph{\JournalTitle{Chaos}}} \textbf{\bibinfo{volume}{25}},
  \bibinfo{pages}{087412} (\bibinfo{year}{2015}).
\newblock \doiprefix 10.1063/1.4928693.

\bibitem{Bleck-02}
\bibinfo{author}{Bleck, R.}
\newblock \bibinfo{journal}{\bibinfo{title}{An oceanic general circulation
  model framed in hybrid isopycnic-{C}artesian coordinates}}.
\newblock {\emph{\JournalTitle{Ocean Modell.}}} \textbf{\bibinfo{volume}{37}},
  \bibinfo{pages}{55--88} (\bibinfo{year}{2002}).

\bibitem{Beron-LaCasce-16}
\bibinfo{author}{Beron-Vera, F.~J.} \& \bibinfo{author}{LaCasce, J.~H.}
\newblock \bibinfo{journal}{\bibinfo{title}{Statistics of simulated and
  observed pair separations in the {Gulf of Mexico}}}.
\newblock {\emph{\JournalTitle{J. Phys. Oceanogr.}}}
  \textbf{\bibinfo{volume}{46}}, \bibinfo{pages}{2183--2199}
  (\bibinfo{year}{2016}).
\newblock \doiprefix 10.1175/JPO-D-15-0127.1.

\bibitem{Beron-Olascoaga-09}
\bibinfo{author}{Beron-Vera, F.~J.} \& \bibinfo{author}{Olascoaga, M.~J.}
\newblock \bibinfo{journal}{\bibinfo{title}{{An assessment of the importance of
  chaotic stirring and turbulent mixing on the West Florida Shelf}}}.
\newblock {\emph{\JournalTitle{J. Phys. Oceanogr.}}}
  \textbf{\bibinfo{volume}{9}}, \bibinfo{pages}{1743--1755}
  (\bibinfo{year}{2009}).
\newblock \doiprefix 10.1175/2009JPO4046.1.

\bibitem{Keating-etal-11}
\bibinfo{author}{Keating, S.~R.}, \bibinfo{author}{Smith, K.~S.} \&
  \bibinfo{author}{Kramer, P.~R.}
\newblock \bibinfo{journal}{\bibinfo{title}{Diagnosing lateral mixing in the
  upper ocean with virtual tracers: Spatial and temporal resolution
  dependence}}.
\newblock {\emph{\JournalTitle{J. Phys. Oceanogr.}}}
  \textbf{\bibinfo{volume}{41}}, \bibinfo{pages}{1512--1534}
  (\bibinfo{year}{2011}).

\bibitem{Cummings-05}
\bibinfo{author}{Cummings, J.~A.}
\newblock \bibinfo{journal}{\bibinfo{title}{Operational multivariate ocean data
  assimilation}}.
\newblock {\emph{\JournalTitle{Q. J. Royal Meteorol. Soc.}}}
  \textbf{\bibinfo{volume}{131}}, \bibinfo{pages}{3583--3604}
  (\bibinfo{year}{2005}).

\bibitem{Cummings-Smedstad-13}
\bibinfo{author}{Cummings, J.~A.} \& \bibinfo{author}{Smedstad, O.~M.}
\newblock \bibinfo{title}{Variational data analysis for the global ocean}.
\newblock In \bibinfo{editor}{Park, S.~K.} \& \bibinfo{editor}{Xu, L.} (eds.)
  \emph{\bibinfo{booktitle}{Data Assimilation for Atmospheric, Oceanic and
  Hydrologic Applications}}, vol.~\bibinfo{volume}{2},
  chap.~\bibinfo{chapter}{13} (\bibinfo{publisher}{Springer-Verlag Berlin
  Heidelberg}, \bibinfo{year}{2013}).
\newblock \doiprefix 10.1007/978-3-642-35088-7-13.

\bibitem{Hadjighasem-etal-13}
\bibinfo{author}{Hadjighasem, A.}, \bibinfo{author}{Farazmand, M.} \&
  \bibinfo{author}{Haller, G.}
\newblock \bibinfo{journal}{\bibinfo{title}{{Detecting invariant manifolds,
  attractors, and generalized KAM tori in aperiodically forced mechanical
  systems}}}.
\newblock {\emph{\JournalTitle{Nonlinear Dyn.}}} \textbf{\bibinfo{volume}{73}},
  \bibinfo{pages}{689--704} (\bibinfo{year}{2013}).

\bibitem{Onu-etal-15}
\bibinfo{author}{Onu, K.}, \bibinfo{author}{Huhn, F.} \&
  \bibinfo{author}{Haller, G.}
\newblock \bibinfo{journal}{\bibinfo{title}{{LCS Tool: A computational platform
  for Lagrangian coherent structures}}}.
\newblock {\emph{\JournalTitle{J. Comp. Sci.}}} \textbf{\bibinfo{volume}{7}},
  \bibinfo{pages}{26--36} (\bibinfo{year}{2015}).

\bibitem{API-NOAA-USCG-EPA2010}
\bibinfo{author}{{National Oceanic and Atmospheric Administration}} \&
  \bibinfo{author}{{U.S. Coast Guard}}.
\newblock \emph{\bibinfo{title}{{Characteristics of Response Strategies: A
  Guide for Spill Response Planning in Marine Environments}}}.
\newblock A joint publication of the {American Petroleum Institute}, the
  {National Oceanic and Atmospheric Administration}, the {U.S. Coast Guard} and
  the {U.S. Environmental Protection Agency} (\bibinfo{year}{2010}).

\bibitem{Melsom2012}
\bibinfo{author}{Melsom, A.}, \bibinfo{author}{Counillon, F.},
  \bibinfo{author}{LaCasce, J.~H.} \& \bibinfo{author}{Bertino, L.}
\newblock \bibinfo{journal}{\bibinfo{title}{{Forecasting search areas using
  ensemble ocean circulation modeling}}}.
\newblock {\emph{\JournalTitle{Ocean Dynamics}}} \textbf{\bibinfo{volume}{62}},
  \bibinfo{pages}{1245--1257} (\bibinfo{year}{2012}).
\newblock \doiprefix 10.1007/s10236-012-0561-5.

\bibitem{Chen2012}
\bibinfo{author}{Chen, C.} \emph{et~al.}
\newblock \bibinfo{journal}{\bibinfo{title}{{FVCOM model estimate of the
  location of Air France 447}}}.
\newblock {\emph{\JournalTitle{Ocean Dynamics}}} \textbf{\bibinfo{volume}{62}},
  \bibinfo{pages}{943--952} (\bibinfo{year}{2012}).
\newblock \doiprefix 10.1007/s10236-012-0537-5.

\bibitem{Lalli-Parsons1993}
\bibinfo{author}{Lalli, C.~M.} \& \bibinfo{author}{Parsons, T.~R.}
\newblock \emph{\bibinfo{title}{{BIOLOGICAL OCEANOGRAPHY: AN INTRODUCTION}}}
  (\bibinfo{publisher}{Pergamon Press Ltd.}, \bibinfo{year}{1993}).

\bibitem{Miller2004}
\bibinfo{author}{Miller, C.}
\newblock \emph{\bibinfo{title}{{BIOLOGICAL OCEANOGRAPHY}}}
  (\bibinfo{publisher}{Blackwell Publishing}, \bibinfo{year}{2004}).

\bibitem{Poje-etal-14}
\bibinfo{author}{Poje, A.~C.} \emph{et~al.}
\newblock \bibinfo{journal}{\bibinfo{title}{{The nature of surface dispersion
  near the Deepwater Horizon oil spill}}}.
\newblock {\emph{\JournalTitle{Proc. Nat. Acad. Sci. USA}}}
  \textbf{\bibinfo{volume}{111}}, \bibinfo{pages}{12693--12698}
  (\bibinfo{year}{2014}).

\bibitem{Jacobs-etal-14}
\bibinfo{author}{Jacobs, G.~A.} \emph{et~al.}
\newblock \bibinfo{journal}{\bibinfo{title}{{Data assimilation considerations
  for improved ocean predictability during the Gulf of Mexico Grand Lagrangian
  Deployment (GLAD)}}}.
\newblock {\emph{\JournalTitle{Ocean Modell.}}} \textbf{\bibinfo{volume}{83}},
  \bibinfo{pages}{98--117} (\bibinfo{year}{2014}).
\newblock \doiprefix 10.1016/j.ocemod.2014.09.003.

\bibitem{Coelho-etal-15}
\bibinfo{author}{Coelho, E.~F.} \emph{et~al.}
\newblock \bibinfo{journal}{\bibinfo{title}{{Ocean current estimation using a
  Multi-Model Ensemble Kalman Filter during the Grand Lagrangian Deployment
  experiment (GLAD)}}}.
\newblock {\emph{\JournalTitle{Ocean Modell.}}} \textbf{\bibinfo{volume}{87}},
  \bibinfo{pages}{86--106} (\bibinfo{year}{2015}).
\newblock \doiprefix 10.1016/j.ocemod.2014.11.001.

\bibitem{Sun2015}
\bibinfo{author}{Sun, S.}, \bibinfo{author}{Hu, C.} \&
  \bibinfo{author}{Tunnell, J.~W.}
\newblock \bibinfo{journal}{\bibinfo{title}{{Surface oil footprint and
  trajectory of the Ixtoc-I oil spill determined from Landsat/MSS and CZCS
  observations}}}.
\newblock {\emph{\JournalTitle{Marine Pollution Bulletin}}}
  \textbf{\bibinfo{volume}{101}}, \bibinfo{pages}{632--641}
  (\bibinfo{year}{2015}).
\newblock \doiprefix 10.1016/j.marpolbul.2015.10.036.

\bibitem{Froyland-etal-07}
\bibinfo{author}{Froyland, G.}, \bibinfo{author}{Padberg, K.},
  \bibinfo{author}{England, M.~H.} \& \bibinfo{author}{Treguier, A.~M.}
\newblock \bibinfo{journal}{\bibinfo{title}{Detection of coherent oceanic
  structures via transfer operators}}.
\newblock {\emph{\JournalTitle{Phys. Rev. Lett.}}}
  \textbf{\bibinfo{volume}{98}}, \bibinfo{pages}{224503}
  (\bibinfo{year}{2007}).

\bibitem{Vukovich2007}
\bibinfo{author}{Vukovich, F.~M.}
\newblock \bibinfo{journal}{\bibinfo{title}{{Climatology of Ocean Features in
  the Gulf of Mexico Using Satellite Remote Sensing Data}}}.
\newblock {\emph{\JournalTitle{Journal of Physical Oceanography}}}
  \textbf{\bibinfo{volume}{37}}, \bibinfo{pages}{689--707}
  (\bibinfo{year}{2007}).
\newblock \urlprefix\url{http://dx.doi.org/10.1175/JPO2989.1}.
\newblock \doiprefix 10.1175/JPO2989.1.

\bibitem{Lindo-Atichati2013}
\bibinfo{author}{Lindo-Atichati, D.}, \bibinfo{author}{Bringas, F.} \&
  \bibinfo{author}{Goni, G.}
\newblock \bibinfo{journal}{\bibinfo{title}{{Loop Current excursions and ring
  detachments during 1993–2009}}}.
\newblock {\emph{\JournalTitle{International Journal of Remote Sensing}}}
  \textbf{\bibinfo{volume}{34}}, \bibinfo{pages}{5042--5053}
  (\bibinfo{year}{2013}).
\newblock
  \urlprefix\url{http://www.tandfonline.com/doi/abs/10.1080/01431161.2013.787504}.
\newblock \doiprefix 10.1080/01431161.2013.787504.

\bibitem{Zavala-Sanson-etal-17}
\bibinfo{author}{{Zavala-Sans\'on}, L.}, \bibinfo{author}{{P\'erez-Brunius},
  P.} \& \bibinfo{author}{Sheinbaum, J.}
\newblock \bibinfo{journal}{\bibinfo{title}{Point source dispersion of surface
  drifters in the southern gulf of mexico}}.
\newblock {\emph{\JournalTitle{Environmental Research Letters}}}
  \textbf{\bibinfo{volume}{12}}, \bibinfo{pages}{024006}
  (\bibinfo{year}{2017}).
\newblock \urlprefix\url{http://stacks.iop.org/1748-9326/12/i=2/a=024006}.

\bibitem{ML-ZH-2009}
\bibinfo{author}{{Mart{\'{\i}}nez-L{\'o}pez}, B.} \&
  \bibinfo{author}{{Zavala-Hidalgo}, J.}
\newblock \bibinfo{journal}{\bibinfo{title}{{Seasonal and interannual
  variability of cross-shelf transports of chlorophyll in the Gulf of
  Mexico}}}.
\newblock {\emph{\JournalTitle{Journal of Marine Systems}}}
  \textbf{\bibinfo{volume}{77}}, \bibinfo{pages}{1--20} (\bibinfo{year}{2009}).
\newblock \doiprefix 10.1016/j.jmarsys.2008.10.002.

\bibitem{Zhang-Hetland-2012}
\bibinfo{author}{Zhang, Z.} \&
  \bibinfo{author}{Hetland, R.}
\newblock \bibinfo{journal}{\bibinfo{title}{{A numerical study on convergence of alongshore flows over the Texas-Louisiana shelf}}}.
\newblock {\emph{\JournalTitle{Journal of Geophysical Research}}}
  \textbf{\bibinfo{volume}{117}(C11010)}, (\bibinfo{year}{2012}).
\newblock \doiprefix 10.1029/2012JC008145.

\bibitem{Goughetal2017}
\bibinfo{author}{Gough, M.} \& \bibinfo{author}{et al.}
\newblock \bibinfo{journal}{\bibinfo{title}{Persistent Lagrangian transport patterns in the northwestern Gulf of Mexico.}}.
\newblock {\emph{\JournalTitle{Submitted JPO}}} 
  (\bibinfo{year}{2017}).
\newblock
  \urlprefix\url{https://arxiv.org/abs/1710.04027}.


\bibitem{LeHenaff2012}
\bibinfo{author}{{Le Henaff}, M.} \emph{et~al.}
\newblock \bibinfo{journal}{\bibinfo{title}{{Surface Evolution of the Deepwater
  Horizon Oil Spill Patch: Combined Effects of Circulation and Wind-Induced
  Drift}}}.
\newblock {\emph{\JournalTitle{Environmental Science and Technology}}}
  \textbf{\bibinfo{volume}{46}}, \bibinfo{pages}{7267--7273}
  (\bibinfo{year}{2012}).

\bibitem{Weisberg2017}
\bibinfo{author}{Weisberg, R.~H.}, \bibinfo{author}{Lianyuan, Z.} \&
  \bibinfo{author}{Liu, Y.}
\newblock \bibinfo{journal}{\bibinfo{title}{{On the movement of Deepwater
  Horizon Oil to northern Gulf beaches}}}.
\newblock {\emph{\JournalTitle{Ocean Modelling}}}
  \textbf{\bibinfo{volume}{111}}, \bibinfo{pages}{81--97}
  (\bibinfo{year}{2017}).
\newblock \doiprefix 10.1016/j.ocemod.2017.02.002.

\bibitem{Barker2011}
\bibinfo{author}{Barker, C.~H.}
\newblock \bibinfo{title}{{A statistical outlook for the Deepwater Horizon oil
  spill}}.
\newblock In \emph{\bibinfo{booktitle}{Monitoring and Modeling the Deepwater
  Horizon Oil Spill: A Record-Breaking Enterprise}}, vol.
  \bibinfo{volume}{195}, \bibinfo{pages}{237--244} (\bibinfo{year}{2011}).
\newblock \urlprefix\url{http://dx.doi.org/10.1029/2011GM001129}.
\newblock \doiprefix 10.1029/2011gm001129.

\bibitem{Ji2011}
\bibinfo{author}{Ji, Z.~G.}, \bibinfo{author}{Johnson, W.~R.} \&
  \bibinfo{author}{Li, Z.}
\newblock \bibinfo{title}{{Oil spill risk analysis model and its application to
  the Deepwater Horizon oil spill using historical current and wind data}}.
\newblock In \emph{\bibinfo{booktitle}{Monitoring and Modeling the Deepwater
  Horizon Oil Spill: A Record-Breaking Enterprise}}, vol.
  \bibinfo{volume}{195}, \bibinfo{pages}{227--236} (\bibinfo{year}{2011}).
\newblock \urlprefix\url{http://dx.doi.org/10.1029/2011GM001117}.
\newblock \doiprefix 10.1029/2011gm001117.

\bibitem{Tulloch2011}
\bibinfo{author}{Tulloch, R.}, \bibinfo{author}{Hill, C.} \&
  \bibinfo{author}{Jahn, O.}
\newblock \bibinfo{title}{{Possible Spreadings of Buoyant Plumes and Local
  Coastline Sensitivities Using Flow Syntheses From 1992 to 2007}}.
\newblock In \emph{\bibinfo{booktitle}{Monitoring and Modeling the Deepwater
  Horizon Oil Spill: A Record Breaking Enterprise}}, \bibinfo{pages}{245--255}
  (\bibinfo{year}{2011}).
\newblock \doiprefix 10.1029/2011GM001125.

\bibitem{Brink2016}
\bibinfo{author}{Brink, K.}
\newblock \bibinfo{journal}{\bibinfo{title}{{Cross-Shelf Exchange}}}.
\newblock {\emph{\JournalTitle{Annual Review of Marine Science}}}
  \textbf{\bibinfo{volume}{8}}, \bibinfo{pages}{59--78} (\bibinfo{year}{2016}).
\newblock
  \urlprefix\url{http://www.annualreviews.org/doi/10.1146/annurev-marine-010814-015717}.
\newblock \doiprefix 10.1146/annurev-marine-010814-015717.

\bibitem{Yang-etal-99}
\bibinfo{author}{Yang, H.}, \bibinfo{author}{Weisberg, R.~H.},
  \bibinfo{author}{Niiler, P.~P.}, \bibinfo{author}{Sturges, W.} \&
  \bibinfo{author}{Johnson, W.}
\newblock \bibinfo{journal}{\bibinfo{title}{Lagrangian circulation and
  forbidden zone on the {West Florida Shelf}}}.
\newblock {\emph{\JournalTitle{Cont. Shelf. Res.}}}
  \textbf{\bibinfo{volume}{19}}, \bibinfo{pages}{1221--1245}
  (\bibinfo{year}{1999}).

\bibitem{Olascoaga-etal-06}
\bibinfo{author}{Olascoaga, M.~J.} \emph{et~al.}
\newblock \bibinfo{journal}{\bibinfo{title}{{Persistent transport barrier on
  the West Florida Shelf}}}.
\newblock {\emph{\JournalTitle{Geophys. Res. Lett.}}}
  \textbf{\bibinfo{volume}{33}}, \bibinfo{pages}{L22603}
  (\bibinfo{year}{2006}).
\newblock \doiprefix 10.1029/2006GL027800.

\bibitem{Olascoaga-NPG-10}
\bibinfo{author}{Olascoaga, M.~J.}
\newblock \bibinfo{journal}{\bibinfo{title}{{Isolation on the West Florida
  Shelf with implications for red tides and pollutant dispersal in the Gulf of
  Mexico}}}.
\newblock {\emph{\JournalTitle{Nonlin. Proc. Geophys.}}}
  \textbf{\bibinfo{volume}{17}}, \bibinfo{pages}{685--696}
  (\bibinfo{year}{2010}).

\bibitem{Sturges2001}
\bibinfo{author}{Sturges, W.}, \bibinfo{author}{Niiler, P.~P.} \&
  \bibinfo{author}{Weisberg, R.~H.}
\newblock \bibinfo{journal}{\bibinfo{title}{{Northeastern Gulf of Mexico Inner
  Shelf Circulation Study}}}.
\newblock {\emph{\JournalTitle{OCS Report MMS. U.S. Minerals Management
  Service}}} \textbf{\bibinfo{volume}{Final Report}}, \bibinfo{pages}{35--1}
  (\bibinfo{year}{2001}).

\bibitem{Li1999a}
\bibinfo{author}{Li, Z.} \& \bibinfo{author}{Weisberg, R.~H.}
\newblock \bibinfo{journal}{\bibinfo{title}{{West Florida shelf response to
  upwelling favorable wind forcing 1: Kinematics}}}.
\newblock {\emph{\JournalTitle{Journal of Geophysical Research}}}
  \textbf{\bibinfo{volume}{104}}, \bibinfo{pages}{13507--13527}
  (\bibinfo{year}{1999}).
\newblock \doiprefix 10.1029/1999JC900073.

\bibitem{Li1999b}
\bibinfo{author}{Li, Z.} \& \bibinfo{author}{Weisberg, R.~H.}
\newblock \bibinfo{journal}{\bibinfo{title}{{West Florida continental shelf
  response to upwelling favorable wind forcing 2. Dynamics}}}.
\newblock {\emph{\JournalTitle{Journal of Geophysical Research}}}
  \textbf{\bibinfo{volume}{104}}, \bibinfo{pages}{23427--23442}
  (\bibinfo{year}{1999}).

\bibitem{ThyngHetland2017}
\bibinfo{author}{Thyng, K.~M.} \& \bibinfo{author}{Hetland, R.~D.}
\newblock \bibinfo{journal}{\bibinfo{title}{Texas and Louisiana coastal
  vulnerability and shelf connectivity}}.
\newblock {\emph{\JournalTitle{Marine Pollution Bulletin}}} 
  (\bibinfo{year}{2017}).
\newblock \doiprefix http://dx.doi.org/10.1016/j.marpolbul.2016.12.074.

\bibitem{Olascoaga2012}
\bibinfo{author}{Olascoaga, M.~J.} \& \bibinfo{author}{Haller, G.}
\newblock \bibinfo{journal}{\bibinfo{title}{{Forecasting sudden changes in
  environmental pollution patterns.}}}
\newblock {\emph{\JournalTitle{Proceedings of the National Academy of Sciences
  of the United States of America}}} \textbf{\bibinfo{volume}{109}},
  \bibinfo{pages}{4738--43} (\bibinfo{year}{2012}).
\newblock \doiprefix 10.1073/pnas.1118574109.


\end{thebibliography}
\end{document}